\documentclass{ws-procs961x669}            
\begin{document}
\title{Galaxy rotation curves and the deceleration parameter in weak gravity}

\author{Maurice H.P.M. van Putten$^*$} 

\address{Sejong University, Seoul 143-747, South Korea\\
$^*$E-mail: mvp@sejong.ac.kr\\
www.sejong.ac.kr}

\begin{abstract}
We present a theory of weak gravity parameterized by a fundamental frequency $\omega_0 = \sqrt{1-q}H$ of the cosmoloogical horizon, where $H$ and $q$ denote the Hubble and, respectively, deceleration parameter. It predicts (i) a $C^0$ onset to weak gravity across accelerations $\alpha = a_{dS}$ in galaxy rotation curves, where $a_{dS}=cH$ denotes the de Sitter acceleration with velocity of light $c$, and (ii) fast evolution $Q(z)=dq(z)/dz$ of the deceleration parameter by $\Lambda=\omega_0^2$ satisfying $Q_0>2.5$, $Q_0=Q(0)$, distinct from $Q_0\lesssim1$ in $\Lambda$CDM. The first is identified in high resolution data of Lellie et al.(2017), the second in heterogeneous data on $H(z)$ over $0<z<2$. A model-independent cubic fit in the second rules out $\Lambda$CDM by $4.35\sigma$ and obtains $H_0=74.0\pm 2.2$ km s$^{-1}$ Mpc$^{-1}$ consistent with Riess et al. (2016). Comments on possible experimental tests by the LISA Pathfinder are included.
\end{abstract}

\keywords{galaxy rotation curves; deceleration parameter; dark energy; dark matter}

\bodymatter

\section{Introduction}

Modern cosmology shows a Universe that is well described by a three-flat Friedmann-Robertson-Walker (FRW) line-element
\begin{eqnarray}
ds^2 = -dt^2 + a(t)^2\left( dx^2 + dy^2 + dz^2 \right),
\label{EQN_FRW}
\end{eqnarray} 
that currently experiences accelerated expansion indicated by a deceleration parameter \cite{rie98,per99}
\begin{eqnarray}
q=\frac{1}{2}\Omega_m - \Omega_\Lambda<0,
\label{EQN_q}
\end{eqnarray}
where $q=-H^{-2}\ddot{a}/a$ with Hubble parameter $H=\dot{a}/a$. It points to a finite dark energy density $\Omega_\Lambda=\rho_\Lambda/\rho_c$, where $\rho_c=3H^2/8\pi G$ is the closure density with Newton's constant $G$. In this background, galaxies and galaxy clusters formed and evolved by weak gravitational attraction at accelerations $\alpha$ at or below the de Sitter scale 
\begin{eqnarray}
a_{dS}=cH,
\label{EQN_adS}
\end{eqnarray}
where $c$ denotes the velocity of light. As such, (\ref{EQN_adS}) sets a scale to {\em weak gravity} common to cosmological evolution and large scale structure.

Weak gravity surprises us with {\em more} than is expected by Newtonian gravitational attraction of observed baryonic matter content: $\Lambda=8\pi\rho_{\Lambda}>0$ and enhanced acceleration in galaxy dynamics. This joint outcome is anticipated by $[\Lambda]=$cm$^{-2}$ and $[\sqrt{\Lambda}]=$cm$^{-1}$ in geometrical units, in which Newton's constant and the velocity of light are set equal to 1, where the latter points to a transition radius $r_t$ in rotation curves of galaxies of mass $M_b$, $4\pi r_t^2\sqrt{\Lambda}\simeq R_g$, $R_g=GM_b/c^2$. Just such scale $r_t$ of a few kpc is observed in a sharp onset to anomalous behavior in high resolution data for $M_b = M_{11}10^{11}M_\odot$ at\cite{lel17,van17}
\begin{eqnarray}
a_N=a_{dS},
\label{EQN_aN}
\end{eqnarray}
where $a_N$ refers to the acceleration anticipated based on Newton's law of gravitational attraction associated with $M_b$ inferred from luminous matter. Further out into weak gravity, these rotation curves satisfy the emperical baryonic Tully-Fisher relation or, equivalently, Milgrom's law.\cite{mil83,mcg11a,mcg11b} 

We here describe a theory of weak gravity by spacetime holography,\cite{bek81,tho93,sus95} parameterized by a fundamental frequency of the cosmological horizon\cite{van17}
\begin{eqnarray}
\omega_0 = \sqrt{1-q}H,
\label{EQN_o0}
\end{eqnarray}
defined by the harmonic oscillator of geodesic separation of null-generators of the cosmological horizon. This result reflects compactness of the holographic phase space of spacetime, set by its finite surface area.. By holography, (\ref{EQN_o0}) induces a dark energy in the 3+1 spacetime (\ref{EQN_FRW}) by the square 
\begin{eqnarray}
\Lambda= \omega_0^2.
\label{EQN_AA}
\end{eqnarray}
 It may be noted that (\ref{EQN_AA}) has vanishing contribution in the radiation dominated era ($q=1$), whereas it reduces to CDM in a matter dominated era ($q=1/2$). (In the equation of state $p=w\rho_\Lambda$ for total pressure $p$, $w=(2q-1)/(1-q)$ vanishes for $q=1/2$.) 
In holographic encoding of particle mass $m_0$ and positions, inertia is defined by a thermodynamic potential $U=mc^2$ derived from unitarity of particle propagators.\cite{van15a,van17} In weak gravity, this may incur \cite{van17}
 \begin{eqnarray}
 m< m_0~~(\alpha < a_{dS})
 \label{EQN_WG}
 \end{eqnarray}
 with a $C^0$ onset to inequality by crossing of apparent Rindler and cosmological horizons. In weak
 gravity, therefore, a particle's inertial mass (``weight") and gravitational mass may appear distinct.\cite{mil99,smo17}

 According to the above, weak gravity prediction the following. First, the $C^0$ onset to (\ref{EQN_WG}) is at a transition radius\cite{van17}
 \begin{eqnarray}
 r_t  = 4.7 \,\mbox{kpc}M_{11}^{1/2}\left(H_0/H\right)^{1/2}.
 \label{EQN_rt}
 \end{eqnarray} 
  Beyond, asymptotic behavior ($\alpha << a_{dS}$) satisfies Milgrom's law with\cite{van16,van17} 
 \begin{eqnarray}
 a_0 = \frac{\omega_0}{2\pi}.
 \label{EQN_a0}
 \end{eqnarray}
 These expressions (\ref{EQN_rt}-\ref{EQN_a0}) may be confronted with data on galaxy rotation curves over an extended redshift range supported by Hubble data $H(z)$.
 Second, the associated dark energy (\ref{EQN_AA}) is relevant to late times, when deceleration $q(z)$ turns negative. At the present epoch, (\ref{EQN_AA}) predicts {\em fast cosmological evolution}, described by $q_0=q(0)$ and $Q_0=Q(0)$, $Q(z)=dq(z)/dz$, satisfying 
 \begin{eqnarray}
  q_0=2q_{0,\Lambda\mbox{CDM}}
  \label{EQN_q0}
  \end{eqnarray}
 and
 \begin{eqnarray}
 Q_0>2.5,~~Q_{0,\Lambda\mbox{CDM}}\lesssim1.
 \label{EQN_Q0}
 \end{eqnarray}

In \S2, we discuss the origin of (\ref{EQN_aN}) in inertia from entanglement entropy and its confrontation with recent high resolution data on galaxy rotation curves.\cite{lel17} In \S3, we discuss (\ref{EQN_AA}) and its implications (\ref{EQN_Q0}) in confrontation with heterogeneous data on $H(z)$ $(0<z<2)$.\cite{sol17,far17} Sensitivity of galaxy dynamics in weak gravity to the Hubble parameter $H(z)$ is studied in \S5.
We summarize our findings in \S6 and conclude with an estimate of $H_0$.

\section{Weak gravity in galaxy rotation curves}

Galaxy dynamics is of particular interest as a playground for the equivalence principle (e.g. \cite{spa16}) at small accelerations on the de Sitter scale $a_{dS}$. 

Einstein's Equivalence Principle (EP) asserts that {\em the gravitational field - wherein photon trajectories appear curved - seen by an observer on the surface of a massive body is indistinguishable from that in an accelerating rocket, at equal weights of its mass}. On this premise, general relativity embeds Rindler trajectories - non-geodesics in Minkowski spacetime - by gravitational attraction as geodesics in curved spacetime around massive bodies, while weight is measured along non-geodesics. With no scale, this embedding is free of surprises, as Rindler accelerations become aribitrarily small. Following such embedding, acceleration vanishes and inertia cancels out in the equations of geodesic motion (ignoring self-gravity). The origin of inertia is hereby not addressed in Einstein's EP or general relativity.

To address inertia for its potential senstivity to a cosmological background, we take one step back and consider Rindler inertia in non-geodesics of Minkowski spacetime.

Inertia is commonly defined by mass-at-infinity in an asymptotically flat spacetime (Mach's principle). The latter is an overly strong assumption in cosmologies (\ref{EQN_FRW}), whose Cauchy surfaces are bounded by a cosmological horizon at Hubble radius $R_H=c/H$. Any reference to large distance asymptotics is inevitably perturbed if not defined by the cosmological horizon. In particular, the apparent horizon $h$ of Rindler spacetime at a distance $\xi=c^2/\alpha$ colludes with the latter at sufficiently small accelerations.
Thus, $h$ and $H$ colludes at (\ref{EQN_aN}) with corresponding transition radius 
\begin{eqnarray}
r_t = \sqrt{R_gR_H},
\label{EQN_A3}
\end{eqnarray}
giving (\ref{EQN_rt}). In what follows, we argue that it sets the onset to {reduced inertia} (\ref{EQN_WG}). 

In what follows, $h$ and $H$ refer to apparent horizons associated with radial accelerations. For orbital motions, we appeal to Newton's separation of particle momenta ${\bf p}$ and associated forces $d{\bf p}/dt$ in radial and azimuthal components, where the former is imparted by the gravitational field of body of mass $M_b$. (Locally, $d{\bf p}/dt$ is measured as curvature of particle trajectories relative to a tangent plane of null-geodesics.) The radial component of $d{\bf p}/dt$ hereby carries $h$ as an apparent horizon defined by instantaneous radial acceleration, giving (\ref{EQN_rt}). In spiral galaxies, (\ref{EQN_A3}) is indeed a typical distance signaling the onset to anamalous galaxy dynamics.\cite{fam12}

\subsection{Origin of Rindler inertia in entanglement entropy}

Unitarity in encoding particle positions by holographic screens satisfies
\begin{eqnarray}
P_-+P_+\equiv1,
\label{EQN_PP}
\end{eqnarray}
where, e.g., $P_\pm$ refer to the probabilites of finding the proverbial cat and, respectively, out of a box, as defined by its quantum mechanical propagator. Satisfying (\ref{EQN_PP}) requires {\em exact arithmetic}, expressing the probability
\begin{eqnarray}
	P_+=1-P_-\simeq 1
\end{eqnarray}
for a cat in the box with no round-off error even when $P_-$ is exponentially small.\cite{van15a} In a holographic approach, $P_+=1-P_-$ is encoded in information {\em on} the box as a compact two-surface with
\begin{eqnarray}
P_- \sim e^{-2\Delta\varphi}
\label{EQN_PM}
\end{eqnarray}
in terms of a Compton phase
\begin{eqnarray}
\Delta\varphi =k\xi,
\end{eqnarray} 
expressing the cat's distance $\xi$ to the walls by Compton wave number $k=mc/\hbar$ for a mass $m$. Exact arithmetic on (\ref{EQN_PP}) requires an information\cite{van15a}
\begin{eqnarray}
I=2\pi \Delta \varphi.
\label{EQN_I}
\end{eqnarray}
Here, the factor of $2\pi$ in (\ref{EQN_I}) is associated with encoding $m$ on a spherical screen. (For a single flat screen, $I=2\Delta\varphi$ and for a cubic box $I=12\Delta\varphi$.) 

We next consider a Rindler observer ${\cal O}$ of mass $m_0$, i.e., a non-geodesic trajectory of constant acceleration in 1+1 Minkowski spacetime $(t,x)$. The light cone at the origin delineates an apparent horizon $h$, $\left|ct\right|=x$, to a Rindler observer with worldline $(t,x)=(\sinh(\lambda\tau),\cosh\lambda\tau)$, where $\lambda = \alpha/c$. ${\cal O}$'s distance $\xi$ to $h$ is Lorentz invariant, which can be attributed an Unruh temperature\cite{unr76}
\begin{eqnarray}
k_BT = \frac{\alpha\hbar}{2\pi c}.
\label{EQN_TU}
\end{eqnarray}
With $h$ null, the entropy $S$, putting Boltzmann's constant $k_B$ equal to 1,
\begin{eqnarray}
dS=-dI
\label{EQN_dS}
\end{eqnarray}
gives rise to a thermodynamic potential
\begin{eqnarray}
dU = -T_UdS = \left(\frac{\hbar\alpha}{2\pi c}\right) \left(2\pi \frac{mc}{\hbar} d\xi \right).
\label{EQN_dU}
\end{eqnarray} 
Conform EP, ${\cal O}$ identifies a uniform gravitational field that it may attribute to some massive object well beyond $h$, featuring a divergent redshift towards $h$. In this gravitational field and relative to $h$, ${\cal O}$ assumes a potential energy
\begin{eqnarray}
U=\int_0^\xi dU= m_0c^2.
\label{EQN_U0}
\end{eqnarray}
By a thermodynamical origin to inertia, Rindler observers experience fluctuations therein equivalently to momentum fluctations by detection of photons from a warm Unruh vacuum.

Emperically, inertia is instantaneus with no associated time scale. Correspondingly, (\ref{EQN_U0}) is not an ordinary thermodynamic potential, but one arising from nonlocal {\em entanglement entropy} $S$ associated with the apparent horizon $h$ in (\ref{EQN_dS}).

\begin{figure}[h]
	\begin{center}
		\includegraphics[width=2.5in]{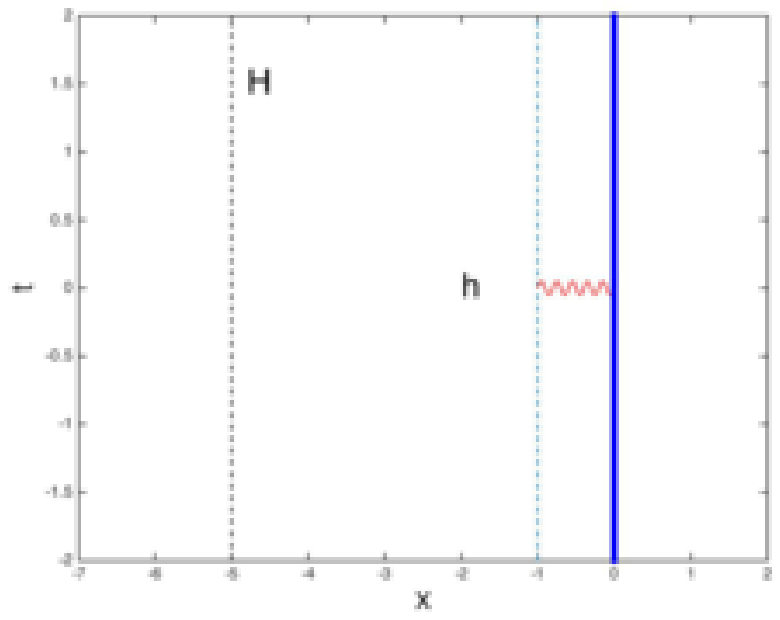}\includegraphics[width=2.7in]{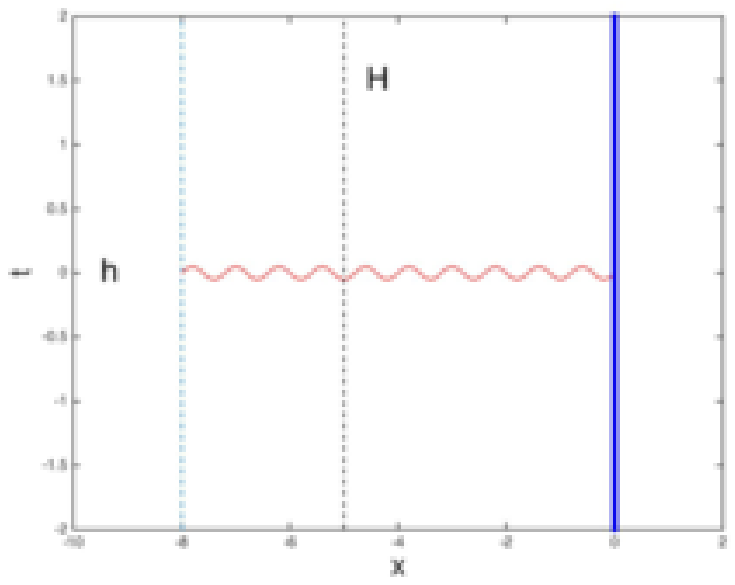}
	\end{center}
	\caption{In three-flat FRW universes with Hubble radius $R_H$, the Rindler horizon $h$ at distance $\xi=c^2/\alpha$ may fall inside (left) or outside the cosmological the cosmological horizon. Collusion at $\alpha=a_{dS}$ of Rindler and cosmological horizon defines a sharp onset to weak gravity. When $H$ falls within $h$, it interferes with the phase space of Rindler observers. Rindler inertia $m$ then drops below its Newtonian value set by rest mass energy $m_0c^2$. (Reprinted from\cite{van17}.)}
	\label{fig1}
\end{figure}

In the absence of any length scale in Minkowksi spacetime, (\ref{EQN_U0}) establishes a Newtonian identity between mass-energy and inertia. In three-flat cosmology, the resulting $m=m_0$ will hold whenever $\alpha>a_{dS}$, ensuring that $h$ falls within $H$ (Fig. \ref{fig1}).

\subsection{$C^0$ onset to weak gravity at $a_{dS}$}

\begin{figure}[h]
	\begin{center}
		\includegraphics[width=3.8in]{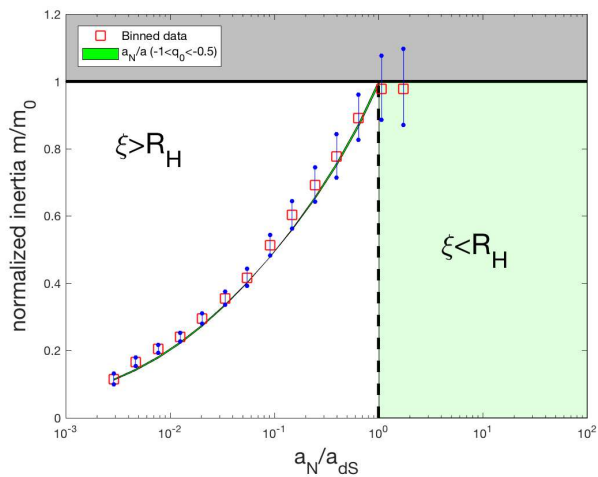}
	\end{center}
	\caption{High resolution data of galaxy rotation curves, here plotted as $m/m_0=a_N/\alpha$ as a function of $a_N/a_{dS}$, where $\alpha$ denotes the observed centripital acceleration and $a_N$ is the expected Newtonian acceleration based on the observed baryonic matter. The results show a sharp onset to weak gravity across $(a_N,m)=(a_{dS},m_0)$. (Reprinted from \cite{van17}.)}
	\label{fig2}
\end{figure}

In (\ref{EQN_WG}), $H$ drops inside $h$ (Fig. \ref{fig1}), and the integral leading to (\ref{EQN_U0}) is cut-off at $H$, leaving $U$ smaller than $m_0c^2$. By $a_{dS}$, EP is no longer scale free, i.e., inertia $m$ measured by $U$ (gravitational pull in an equivalent gravitational field) may deviate from the Newtonian value $m_0$. 
Reduced inertia $m<m_0$ when $h$ falls beyond $H$ in (\ref{EQN_WG}) gives rise to enhanced acceleration at a given Newtonian gravitational forcing $F_N=m_0M_b/r^2$. (This is prior to a covariant embedding in geodesic motion in curved spacetime, alluded to above.) 
Arising from crossing of the apparent Rinder and cosmological horizon surfaces, the onset to $m< m_0$ is $C^0$ sharp. Just such behavior is apparent in high resolution galaxy rotation data (Fig. \ref{fig2}).

Fig. \ref{fig2} shows binned rotation curve data from many spiral galaxies, after normalizing the independent variable to $a_N$ (rather than the coordinate radius $r$).\cite{fam12} Plotting $\alpha/a_N$ as a function of $a_N$ (or $a_N/a_{dS}$), the results $\alpha>a_N$ are often referred to as the ``missing mass problem." In light of our focus on Rindler horizons falling beyond the cosmological horizon, Fig. \ref{fig2} shows 
\begin{eqnarray}
	\frac{m}{m_0}=\frac{a_N}{\alpha}.
\label{EQN_in}
\end{eqnarray}
Observationally, the right hand side of (\ref{EQN_in}) is inferred from the ratio $(V^2/r)/(V_b^2/r)=(V/V_b)^2$ of observed $(V)$ and anticipated $(V_c)$ circular velocities, the latter based on luminous matter. Hence,
\begin{eqnarray}
\frac{E_k}{E_{k,0}} = \frac{mV^2}{m_0V_b^2} = \frac{m\alpha}{m_0a_N} =1,
\end{eqnarray}
showing that orbital kinetic energies are unchanged.

In weak gravity (\ref{EQN_WG}), holographic representations of a particle of mass $m$ involves two low energy dispersion relations of image modes in 3+1 spacetime and of Planck sized surface elements on the cosmological horizon, satisfying\cite{van16,van17}
\begin{eqnarray} 
\hbar\omega = \sqrt{\hbar^2\Lambda + c^2p^2} ~~(r<R_H),~~T=\sqrt{T_U^2+T_{H}^2}~~(r=R_H),
\label{EQN_DR1}
\end{eqnarray}
where $k_BT_{H}=\hbar a_{H}/(2\pi c)$ denotes the cosmological horizon temperature at internal surface gravity 
$a_H = (1/2)(1-q)H$. At equal mode counts, we have\cite{van17}
\begin{eqnarray}
2B(p)\frac{m}{m_0} = \frac{R_H}{\xi}
\label{EQN_DR2}
\end{eqnarray}
with the momentum dependent ratio 
\begin{eqnarray}
B(p)=\frac{\sqrt{\hbar^2\Lambda + c^2p^2}}{k_B\sqrt{T_U^2+T_H^2}}.
\label{EQN_Bp}
\end{eqnarray}
Rindler's relation $\xi = c^2/\alpha$ hereby obtains
\begin{eqnarray}
\alpha = \sqrt{2B(p)a_{dS}a_N},
\label{EQN_alpha1}
\end{eqnarray}
where $a_N$ refers to the Newtonian acceleration $a_N = M_b/r^2$. An effective description of weak gravity (\ref{EQN_WG}) 
now follows, upon taking a momentum average $\left < B(p)\right>$ of (\ref{EQN_Bp}) in (\ref{EQN_alpha1}). Averaging over a thermal distribution \cite{van17} obtains the green curve in Fig. \ref{fig2}.

In the asymptotic regime, $a_N<<a_{dS}$, (\ref{EQN_alpha1}) reduces to Milgrom's law\cite{mil83,fam12} 
\begin{eqnarray}
\alpha = \sqrt{a_0 a_N}
\label{EQN_alpha2}
\end{eqnarray}
with Milgrom's parameter (\ref{EQN_a0}) directly asociated with the background cosmology as anticipated based on dimensional analysis in geometrical units.\cite{van16,van17}
In light of the $q$-gradients $Q_0$ in (\ref{EQN_Q0}), $A_0=A(0)$, $A(z)=a_0^{-1}da_0(z)/dz$, satisfies\cite{van16,van17}
\begin{eqnarray}
A_{0}\simeq -0.5,~~A_{0,\Lambda\mbox{CDM}}>0.
\label{EQN_A0}
\end{eqnarray} 
This discrepant outcome of (\ref{EQN_AA}) and $\Lambda$CDM may be tested observationally in surveys of rotation curves covering a finite redshift range.

\section{Accelerated expansion from cosmologial holography}

The holographic principle proposes a reduced phase space of spacetime, matter and fields to that of a bounding surface in Planck sized degrees of freedom.\cite{bek81,tho93,sus95} As the latter is generally astronomically large in number, and hence excited at commensurably low energies. 
In weak gravity, these low energies readily reach a low energy scale of the cosmological vacuum, set by the 
cosmological horizon at Hubble radius $R_H={c}/{H}$ based on the unit of luminosity $L_0=c^5/G$.\cite{van15a} 

As a null-sphere, the cosmological horizon features closed null-geodesics. The geodesic deviation of a pair of null-geodesics satisfies a harmonic oscillator equation\cite{van17}. By explicit calculation of the Riemann tensor in a three-flat FRW universe with deceleration parameter $q$, the angular frequency satisfies (\ref{EQN_o0}). A holographic extension to spacetime within obtains a wave equation with dispersion relation $\omega = \sqrt{k^2+\Lambda}$, where $k$ refers to the wave number of modes orthogonal to the cosmological horizon and $\Lambda$ in (\ref{EQN_AA}). Dark energy is hereby described in terms of the canonical cosmological parameters $(H,q)$, allowing a detailed comparison of cosmic evolution with (\ref{EQN_AA}) versus $\Lambda$CDM.

\subsection{Evolution by $\Lambda=\omega_0^2$ and $\Lambda$CDM}

\begin{figure}[h]
\begin{center}
 \includegraphics[width=4.0in]{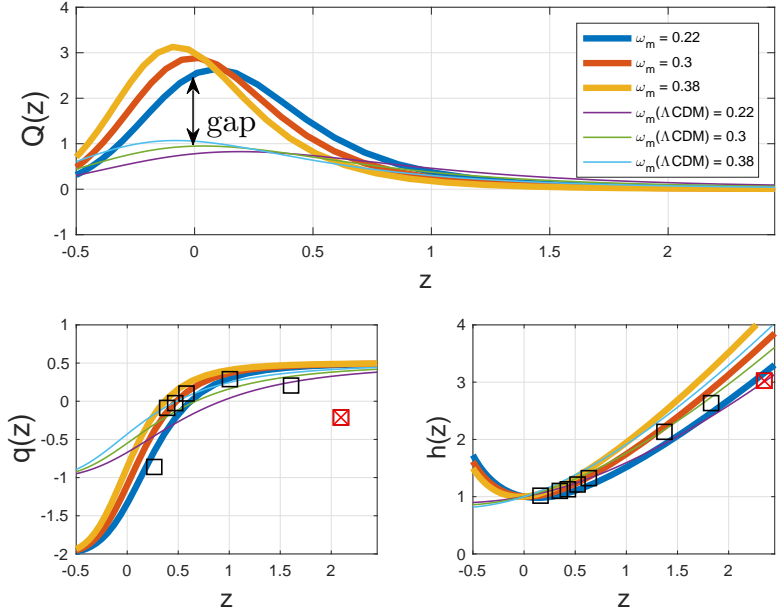}
\end{center}
\caption{Evolution of the deceleration parameter $q(z)$ with redshift in a three-flat FRW universe with dark energy $\Lambda=\omega_0^2$ and $\Lambda$CDM for various values $q_0=q(0)$. They show pronounced differences in slope $Q_0$ at $z=0$ associated with different curvatures of $h(z)=H(z)/H_0$.}
\label{FRW}
\end{figure}

FRW universes with $\Lambda$ in (\ref{EQN_AA}) leave baryon nucleosynthesis invariant, since $q=1$ in a radiated dominated epoch. For most of the subsequent evolution of the universe, this dark energy is relatively small compared to matter content. In recent epochs, however, the derivative 
\begin{eqnarray}
Q(z) = \frac{dq(z)}{dz}
\end{eqnarray} 
varies rather strongly compared to what is expected in $\Lambda$CDM, expressed in (\ref{EQN_Q0}) and 
shown in Fig. \ref{FRW}. 

Cosmological evolution in the approximation of (\ref{EQN_FRW}) is described by an ordinary differential equation for the scale factor $a(t)$. For (\ref{EQN_AA}), this is described by an ordinary differential equation (ODE) as a function for $y=\log h$, $h=H/H_0$, as a function of $z$,
\begin{eqnarray} 
y^\prime(z)= 3(1+z)^2\omega_m e^{-2y} - (1+z)^{-1}
\end{eqnarray} 
derived from $\ddot{a}(\tau)= a\left( 2h^2 - 3\omega_m a^{-3}  \right)$ ($a(0)=1$, $h(0)=1$),
$d/dt = - (1+z)Hd/dz$, as a function of time $\tau = tH_0$.\cite{van15b}
For $\Lambda$CDM, we have $h(z) = \sqrt{(1-\omega_m)+\omega_m(1+z)^{3}}$ for $\Lambda$CDM.
Fig. 2 shows illustrative numerical solutions for various values $\omega_m$ of dark matter content at $z=0$.
At late times, the dynamical dark energy (\ref{EQN_AA}) features a {\em fast} expansion compared to $\Lambda$CDM. In particular, $q_0$ and $Q_0$ defined in (\ref{EQN_q0}-\ref{EQN_Q0}) are markedly distinct in these two models. 

\begin{table}
	\tbl{Binned data $\{z_k,H(z_k)\}$ $(k=1,2,\cdots 8)$\cite{far17} on the Hubble parameter $H(z)$
		$\left[\mbox{km~s}^{-1}\mbox{Mpc}^{-1}\right]$ over an extended range of redshiftz $z$, and
		inferred estimates of $H^\prime(z_{k^\prime})$ and $q(z_{k^\prime})$ at 
		midpoints $z_{k^\prime} = (z_k+z_{k-1})/2$ $(k=2,3,\cdots,8)$, and $Q(z_k)$ $(k=2,3,\cdots,7)$.}
	{\begin{tabular}{@{}lccccccc@{}}
			\toprule
			$k$ &  redshift $z_k$, $z_{k^\prime}$ & $H(z_k)$ & $\sigma_k$ & $H^\prime(z_{k^\prime})$ & $q(z_{k^\prime})$ & $Q(z_{k})$ \\
			\colrule
			1 & 0.166 & 75.7 & 3.35 &     \\
			& 0.2605 & 78.20 & - &  26.46 & -0.5736 \\
			2& 0.355 & 80.7 & 1.70 &   -       & -           & 1.9758\\
			& 0.3910 & 82.65 & - & 54.17 & -0.0884\\
			3 & 0.427 & 84.6 & 4.80 & -       & -           & 0.2659\\
			& 0.4725 & 87.35 & - & 60.44 & -0.0189 \\
			4 & 0.518 & 90.1 &  1.75 & -       & -           &  0.1492 \\
			& 0.5755 & 93.90 & - & 66.09 & 0.1008 \\
			5 & 0.633 & 97.7 & 1.90 & -        & -          &  0.0913 \\ 
			& 1.0015 & 127.85 & - & 81.82 & 0.2809 \\
			6 & 1.37    & 158 &  8.50 & -       & -          & -0.1470\\ 
			& 1.60 & 177 & - & 82.61 &  0.2135\\ 
			7 & 1.83    & 196 & 31.0  & -       & -          & -0.2328\\
			& 2.09 & 210 &  -  & 53.85 & -0.2077 \\
			8 & 2.35   & 224 & 5.0    &  \\
			\botrule
		\end{tabular}
	}
	\label{Table_val}
\end{table}

\subsection{Heterogeneous data on $H(z)$}

Measurements of the Hubble parameter $H(z)$ by various methods of observations now extends over an increasingly large redshift range. For recent compilations, see, e.g., Sola et al.\cite{sol17} covering  $0<z < 1.936$ and Farooq et al.\cite{far17} covering $0<z\le 2.36$. 

Table \ref{Table_val} shows $N=8$ binned data\cite{far17} on $H(z)$ with a mean of normalized standard deviation $\hat{\sigma}_k=\sigma_k/H(z_k)$ satisfying
\begin{eqnarray} 
\left\{ \frac{1}{N}\sum_{k=1}^N \hat{\sigma}_k^{-2}\right\}^{-\frac{1}{2}} \simeq 3\%.
\end{eqnarray}
Included are the estimates of $q(z)$ obtained from $H^\prime(z)$ by central differencing.

As heterogeneous data sets, these compilations require tests against physical constraints before they can be used in regression analysis. Different (often unknown) systematics can potentially create trend anomalies that violate essential priors. If so, the data set contains {\em incompatibilities}.

We recall that the Universe entered an essentially matter dominated epoch when $z$ appreciably exceeds the transition redshift $z_t$ defined by the vanishing of the deceleration paramer $q(z)$, when the Hubble flow of galaxies passing through the cosmological horizon changed sign. The constraint $z_t<1$ seems fairly robust\cite{far17}. For $z>1$, therefore, the positivity conditions
\begin{eqnarray}
H(z)>0,~~q(z) = -1 + (1+z)H^{-1}(z)H^\prime(z) > 0
\label{EQN_qp}
\end{eqnarray}
are {\em essential physical priors} to any data set.

Table 1 points to a violation of (\ref{EQN_qp}) with $q(z)=-0.2077$ at the midpoint $z=2.09$. Associated with a drop in $H^\prime(z)$, the last data point at $k=8$ appears to be incompatible with $k=1,2,\cdots,7$ and will be considered incompatble in the following nonlocal analysis.

\subsection{Detecting $Q_0$ in $H(z)$ data}

\begin{figure}[h]
 	\begin{center}
 		\includegraphics[width=5.0in]{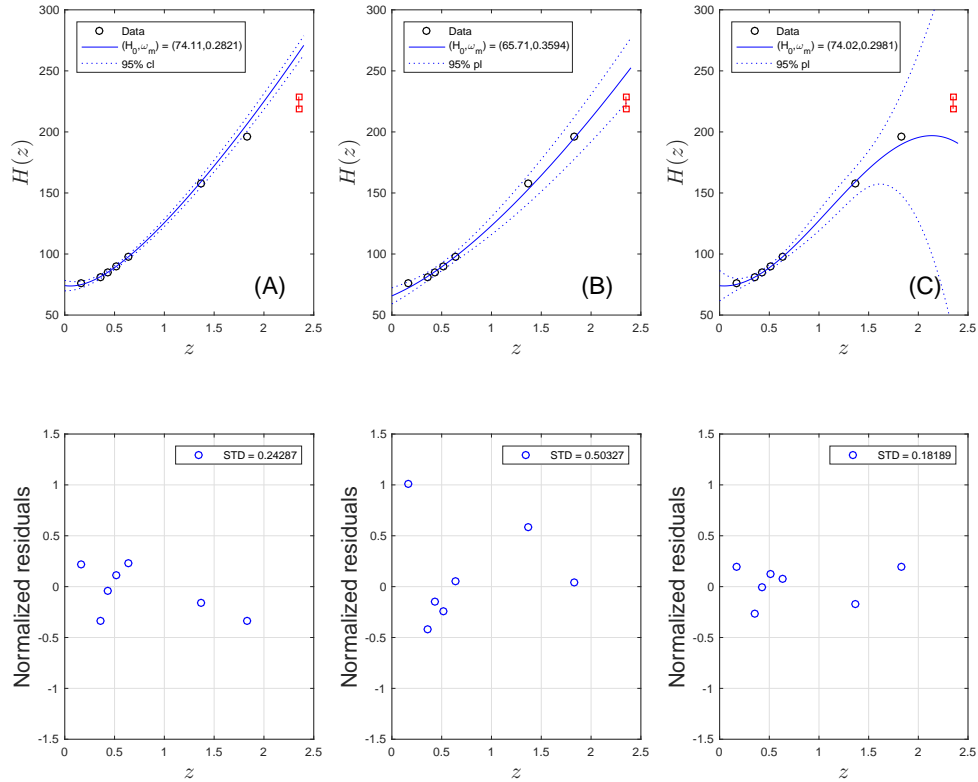}
 	\end{center}
 	\caption{Nonlinear regression model fits (continuous line) to the binned data (circles) of Table \ref{Table_val} $(k=1,2,\cdots,7)$ by the MatLab function {\em fitnlm} to (A) $\Lambda=\omega_0^2$, (B) $\Lambda$CDM, (C) a cubic polynomial. Included are the 95\% prediction limits (pl, dashed lines; very similar to the 95\% confidence limits, not shown). Included is the incompatible data point $k=8$ (red) in Table \ref{Table_val}.}
 	\label{Qz1}
 \end{figure}

Table \ref{Table_res} lists results of nonlinear model regression by our two cosmological models to the Hubble data of Table \ref{Table_val}, parameterized by $(H_0,\omega_m)$ with the MatLab function {\em fitnlm} using weights $w_k\propto \sigma_k^{-2}$.\cite{pod01,far17} Fig. \ref{Qz1} shows the fits and 95\% prediction limits (pl, very close but slightly wider than 95\% confidence limits). 

Fig. \ref{Qz1} shows the match of evolution with (\ref{EQN_AA}) including the normalized errors, further for $\Lambda$CDM and a model-independent cubic fit. 

For (\ref{EQN_AA}) these errors are essentially the same as obtained by a model-independent cubic fit
\begin{eqnarray}
H(z) = H_0 \left(1 + (1+q_0) z +\frac{1}{2}\left(Q_0+(1+q_0)q_0\right)z^2+b_4z^3\right) + O\left(z^4\right)
\label{EQN_hC}
\end{eqnarray}
with free coefficients $(H_0,q_0,Q_0,b_4)$, and they are about one-half the normalized errors in the fit by $\Lambda$CDM.

For the two models, $q_0$ and $Q_0$ obtain according to their repective evolution equations, i.e.,
$\Omega_M=(1/3)(2+q_0)$ for (\ref{EQN_AA}), (\ref{EQN_q0}) for $\Lambda$CDM, and
\begin{eqnarray}
Q_0 = 
\left\{\begin{array}{ll}
(2+q_0)(1-2q_0)~>~2.5 & (\Lambda = \omega_0^2),\\ \\
(1+q_0)(1-2q_0)~\lesssim~1 &  (\Lambda\mbox{CDM}),
\end{array}\right.
\label{EQN_Q00}
\end{eqnarray}
whereas $Q_0$ in our cubic fit (\ref{EQN_hC}) is determined directly by nonlinear regression.

Included in Fig. \ref{Qz1}) is the $k=8$ data point in Table \ref{Table_val}. As expected, it is inconsistent with (\ref{EQN_AA}) and only marginally consistent with $\Lambda$CDM.
 \begin{table}
 	\tbl{Results of nonlinear model regression on the binned compatible data $\{z_k,H(z_k)\}$ $(k=1,2,\cdots 7)$ of Table \ref{Table_val}.}
	{\begin{tabular}{@{}lccccccc@{}}
 			\toprule
 			model &  $H_0$ $\left[\mbox{km~s}^{-1}\mbox{Mpc}^{-1}\right]$ & $\omega_m$ & $q_0$ & $Q_0$\\
 			\colrule
 			$\Lambda=\omega_0^2$ & $74.1\pm1.2$  & $0.2821\pm0.0125$  & $-1.1537\pm0.0375$ & $2.7990\pm0.0603$  \\
 			$\Lambda$CDM         & $65.7\pm 1.7$ & $0.3594\pm0.0384$      & $-0.4609\pm0.0576$ & $1.0360\pm0.0486$  \\
 			cubic fit                     & $74.0\pm2.4$ & $0.2981\pm0.0566$      & $-1.1057\pm0.1698$ & $2.2648\pm0.2910$  \\
 			\botrule
 		\end{tabular}}
 	\label{Table_res}
 \end{table}

In Table 2, $Q_0=2.2648\pm0.2910$ of the model-independent cubic fit rules out $\Lambda$CDM according to (\ref{EQN_Q00}) by $4.36\,\sigma$.

\section{Sensitivity to $H(z)$ in high redshift galaxy rotation curves}

By continuity in the onset to weak gravity at $\alpha=a_{dS}$ (\ref{EQN_rt}), we have 
\begin{eqnarray}
y_{0,h}=\left(\frac{a_N}{a_{dS}}\right)_h =\left( \frac{r_t}{R_h}\right)^2
\end{eqnarray}
whereby (\ref{EQN_alpha1}) takes the form
\begin{eqnarray}
\frac{\alpha}{a_{dS}} = \sqrt{\mu} \frac{r_t}{R_h}
\label{EQN_alpha3}
\end{eqnarray}
with $\mu = 2\left<B(p)\right>$. In weak gravity $\alpha \lesssim a_{dS}$, anomalous behavior in galactic dynamics may be expressed by an apparent equivalent dark matter fraction
\begin{eqnarray}
f_{DM}^\prime = \frac{\alpha - a_N}{\alpha},
\label{EQN_fDM}
\end{eqnarray}
conform the definition of $f_{DM}$ in\cite{gen17}.

\begin{figure}[h]
			\begin{center}
				\includegraphics[width=5.0in]{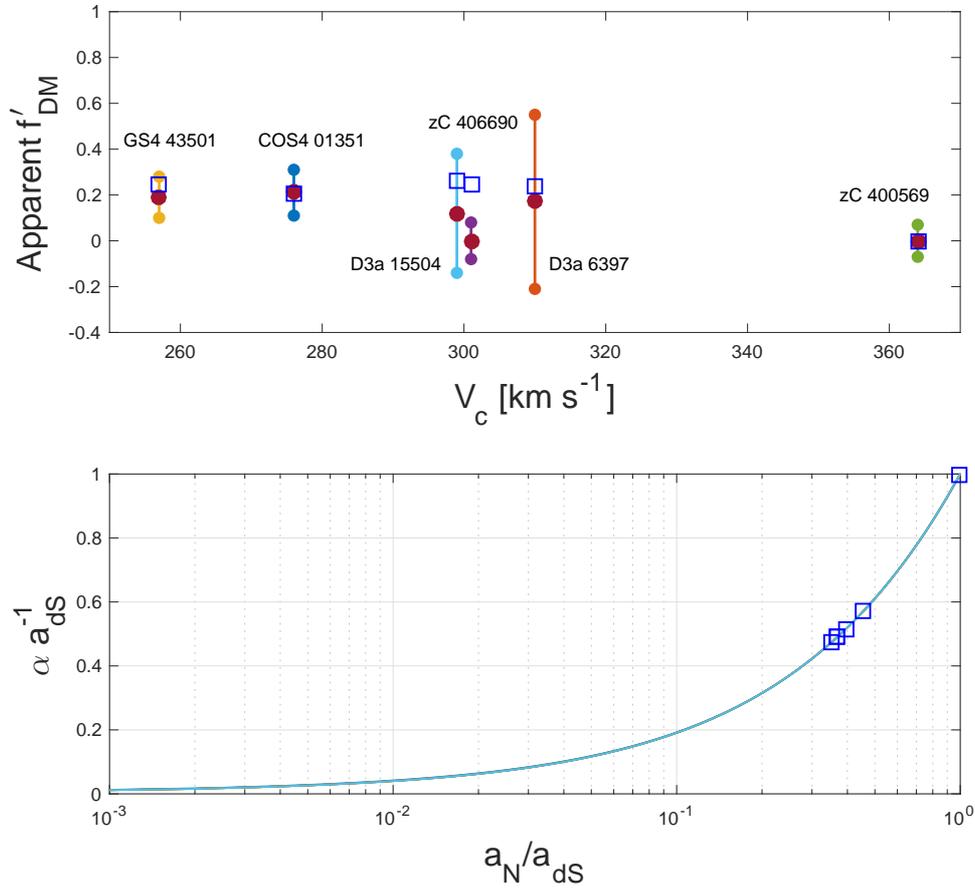}
			\end{center}
			\caption{(Upper panel.) Apparent dark matter fractions $f_{DM}^\prime$ (blue squares) and observational data on $f_{DM}$. Results agree except for cZ 406690 ($V_c=301$ km/s). (Lower panel.) Location of the data in Table 3 in weak gravity $a_N<a_{dS}$ and associated predicted accelerations $\alpha/a_{dS}$. (Reprinted from\cite{van17c}.) }
			\label{figG}
		\end{figure}

Table 3 lists data on high redshift sample of rotation curves\cite{gen17} and associated data on  (\ref{EQN_alpha2}-\ref{EQN_fDM}). Fig. \ref{figG} shows the apparent ($f_{DM}$) and predicted ($f_{DM}^\prime$) dark matter fractions. Based on $r_t/R_H$, this sample of galaxies probes weak gravity (\ref{EQN_alpha1}) in $\alpha < a_{dS}$ but not the asymptotic regime $\alpha << a_{dS}$, a point recently empasized by\cite{mil17}.

\begin{table}[h]
	\tiny 
	{\bf Table 3.} {Analysis on apparent $(f_{DM}^\prime)$ versus observed $(f_{DM})$ dark matter fractions in high $z$ rotation curves\cite{gen17} with baryonic mass $M_b=M_{11}10^{11}M_\odot$ and rotation velocities $V_c$ in units of km\,s$^{-1}$ at the half-light radius $R_h$.}
	\centerline{	{\begin{tabular}{@{}lccccccc|cl@{}}
				\mbox{}\\\hline\hline
				Galaxy &  $z$ & $H(z)/H_0$ & $q(z)$ & $M_{11}$ & $r_t/R_h$ & $\mu$ & $f^\prime_{DM}$ & $V_c$ & $f_{DM}$ (95\% c.l.)\\
				\hline
				COS4 01351 & 0.854  & 1.5986 & 0.0853 &  1.7  & 0.6740 & 0.7050 & 0.2031 &276 & 0.21$\pm0.1$\\
				D3a 6397 & 1.500  & 2.2883 & 0.2957 & 2.3  & 0.6273 & 0.6587 & 0.2342 & 310 & 0.17 $(<0.38)$\\
				GS4 43501 & 1.613  & 2.4246 & 0.3170 &  1.0  & 0.6054 & 0.6381 & 0.2497 & 257 & 0.19 $(\pm0.09)$\\
				zC 406690 & 2.196  & 3.1903 & 0.3996 & 1.7  & 0.6060 &0.6383 & 0.2489 & 301 & 0 $(<0.08)$\\
				zC 400569 & 2.242 & 3.2531 & 0.3919 &  1.7  & 0.9992 & 0.9814 & 0.0006 & 364 & 0 $(<0.07)$\\ 
				D3a 15504 & 2.383 & 3.4537 & 0.4184 &  2.1 & 0.5908 & 0.6239 & 0.2590 & 299 & 0.12 $(<0.26)$\\
			\end{tabular}}}
			\label{Table_val3}
		\end{table}
		
		Fig. \ref{figG} shows quantitative agreement of $f_{DM}$ and $f_{DM}^\prime$, except for cZ 4006690 ($V_c=301$ km s$^{-1}$). We note that its observed rotation curve is asymmetric, which may indicate systematic errors unique to this galaxy. (The other galaxies all have essentially symmetric rotation curves.) Apart from this particular galaxy, $f_{DM}^\prime\simeq f_{DM}$ in over a broad range of redshifts confirms that (\ref{EQN_rt}), and hence weak gravity in galaxy dynamics is co-evolving with $H(z)$ in the background cosmology.

 \section{Conclusions and outlook}
 
By volume, structure formation and cosmology represent the most common gravitational interactions at the scale of $a_{dS}$ or less. Relatively strong interactions in our solar system and its general relativistic extension to compact objects are, in fact, the exception. A principle distinction between the two is sensitivity to $a_{dS}$ in weak gravitation, otherwise ignorable in the  second. This potential is completely absent in Einstein's EP and classical formulations of general relativity, but becomes a real possibility by causality and compactness of the cosmological horizon.

Inspired by spacetime holography, weak gravity in galactic dynamics and cosmological evolution is parameterized by $a_{dS}$ and the fundamental requency $\omega_0$ of the cosmological horizon, set by the canonical parameters $(H,q)$ in (\ref{EQN_FRW}). Observational consequences are (i) a perturbed inertia at small accelerations $(\alpha < a_{dS})$ across a sharp onset at $a_N = a_{dS}$ with asymptic behavior ($a_N<<a_{dS})$ described by a Milgrom's parameter $a_0 = \omega_0/2\pi$; and (ii) $Q_0>2.5$ at accelerated expansion by $\Lambda = \omega_0^2$.

Our $\Lambda=\omega_0^2$ is completely independent of the bare cosmological constant $\Lambda_0$ arising from vacuum fluctuations in quantum field theory. Complementary to unitary holography based on particle propagators, vacuum flucations with no entanglement have a trivial Poincare invariant propagator. By translation invariance, vacuum fluctuatons do not drag spacetime like matter does, whereby $\Lambda_0$ has no inertia and carries no gravitational field conform Mach's principle.

Predictions (i-ii) are supported by the results of Fig. 1 and Table 2:
\begin{itemize} 
\item Rotation curve data\cite{lel17} point to a $C^0$ onset to weak gravity in Fig. 1 at 
\begin{eqnarray}
\left( \frac{a_N}{a_{dS}},\frac{\alpha}{a_N}\right)=(1,1).
\end{eqnarray}
We attribute this behavior to inertia originating in entanglement entropy, of the apparent Rindler or cosmological horizon, whichever is more nearby. At small accelerations, inertia is reduced when $h$ falls beyond $H$, enhancing acceleration for a given gravitational forcing. By (\ref{EQN_rt}), (\ref{EQN_a0}) and (\ref{EQN_A0}), weak gravity is sensitive to background cosmology, manifest in galaxy rotation curves over an extended redshift range listed in Table 3 and shown in Fig. \ref{figG}.
\item 
A model-independent analysis of the latter by a cubic polynonial identifies $Q_0\simeq 2.26\pm0.29$. $\Lambda$CDM with $Q_0\lesssim1$ is hereby ruled out at $4.36\,\sigma$. Normalized errors in the cubic fit and those to $\Lambda=\omega_0^2$ are about one-half of the normalized errors in the fit to $\Lambda$CDM. 
\item
Estimates of $H_0$ in the first two are consistent with recent measurement $H_0 = 73.24\pm 1.74$\,km\,s$^{-1}$Mpc$^{-1}$ in local surveys\cite{rie16,li16}. Combining our result for $\Lambda = \omega_0^2$ with the latter obtains relatively fast expansion compared to $\Lambda$CDM with
\begin{eqnarray}
H_0=73.8 \pm 0.9\,\mbox{km}\,\mbox{s}^{-1}\,\mbox{Mpc}.
\label{EQN_REF2}
\end{eqnarray}
\end{itemize}

The above suggests a strong form of the EP, by inisting on equivalent fluctuations in inertia of Rindler observers (non-geodesics in Minkowski spacetime) with fluctuations in measuring weight on a scale (non-geodesics in curved spacetime). Momentum fluctuations between two bodies in mutual gravitational attraction herein are fully entangled, ensuring preservation of total momentum. If so, curvature from mass must follow very similar rules giving rise to (\ref{EQN_U0}). Consequently, any scenario for entropic gravity\cite{ver11} should include the origin of inertia in entanglement entropy, the details of which remain to be spelled out.

Recently, the LISA Pathfinder has conducted an extensive set of a high precision free fall gravity and inertia experiments at accelerations well below $a_{dS}$.\cite{dan17} It uses laser-interferometry and electrostatic forcing to track and perturb test masses about their geodesic motion, in the spacecraft's self-gravitational field. In probing the de Sitter scale of 1\,\AA\,s$^{-2}$ or less, we note that (\ref{EQN_rt}) conventiently rescales to the size of a small spacecraft, i.e.,
\begin{eqnarray}
	r_t \simeq 30\,\mbox{cm}\,m_{1}^{1/2}
	\label{EQN_A4}
\end{eqnarray}
about a gravitating mass $m=m_1$\,kg, representative for the 2 kg test masses in the LISA Pathfinder mission. It seems worthwhile to look at data covering $\alpha< a_N$. 

{\bf Acknowledgements.} The author thanks stimulating discussions with S.S. McGaugh, J. Binney, K. Danzmann, T. Piran, G. Smoot, C. Rubbia,  J. Sol\`a, L. Smolin and M. Milgrom. This report was supported in part by the National Research Foundation of Korea under grant No. 2015R1D1A1A01059793 and 2016R1A5A1013277.

\end{document}